\documentclass[twocolumn,preprintnumbers,amsmath,amssymb,final]{revtex4}

\usepackage{graphicx}
\usepackage{dcolumn}
\usepackage{bm}

\def\BEq{\begin{equation}}
\def\EEq{\end{equation}}
\def\BEqA{\begin{eqnarray}}
\def\EEqA{\end{eqnarray}}
\def\BEn{\begin{enumerate}}
\def\EEn{\end{enumerate}}
\def\BWT{\begin{widetext}}
\def\EWT{\end{widetext}}

\def\bra{\langle}

\def\ket{\rangle}

\begin{document}


\title{
Simple protocol for generating W states
in resonator-based quantum computing architectures
}

\author{Andrei Galiautdinov}
\email{ag@physast.uga.edu}
\affiliation{
Department of Physics and Astronomy,
University of Georgia, Athens, Georgia
30602, USA
}

\date{\today}

\begin{abstract}

 We describe a simple, practical scheme for generating
multi-qubit W states in resonator-based architectures, in which $N$ 
Josephson phase qubits are
capacitively coupled to a common resonator bus. 
The entire control sequence consists of three pulses: a local Rabi pulse that excites a 
single qubit in the circuit; a coupling pulse that transfers the qubit excitation to the 
resonator bus; and the main, entangling operation that simultaneously couples the bus 
to all $N$ qubits. 
If the qubit-resonator coupling strength $g$ is much smaller 
than the qubit energy splitting $E_{10}$, the system initially excited into the 
near-degenerate single-excitation subspace stays within that subspace, while smoothly 
evolving toward the 
fully uniform W state superposition. 
The duration of the final entangling operation is found to {\it decrease} with the total 
number of the qubits according to $t = \pi/(2g\sqrt{N})$, in agreement with some of the 
previously proposed cavity QED W state generation schemes.

\end{abstract}


\maketitle

%

\section{Description of the W protocol}

Our control sequence consists of the following three steps:
\begin{enumerate}
\item
First, the initial local Rabi pulse is applied to one
of the qubits in the circuit, bringing the qubit from its ground state $|0\ket$ to 
the excited state 
$|1\ket$,
\BEq
|00\dots 000_{r}\ket \rightarrow |00\dots 010_{r}\ket.
\EEq
\item
The corresponding qubit-resonator coupling is then turned on, which transfers 
the excitation 
state from the qubit to the bus,
\BEq
|00\dots 010_{r}\ket \rightarrow |00\dots 001_{r}\ket.
\EEq
\item
The second entangling pulse is applied, which couples 
the bus to all the qubits in the system. If the coupling $g$ is much smaller 
than the qubit energy splitting $E_{10}$, the system initially prepared in the 
single-excitation subspace stays within that subspace while smoothly evolving 
toward the 
multi-qubit, fully uniform W-state superposition as follows,
\BEqA
&&|00\dots 001_{r}\ket \rightarrow
\nonumber \\
&&
\left[
\frac{|00\dots 01\ket + |00\dots 10\ket
+\dots + |10\dots 00\ket}{\sqrt{N}}
\right]\otimes |0_{r}\ket.
\nonumber \\
\EEqA
This last step is similar to the W state generation scheme proposed for cavity 
QED in Ref.\ \cite{MingYang2004}.
 In the terminology of reference \cite{MIGLIORE2006}, 
our resonator bus plays the role of the entanglement mediator.

\end{enumerate}

\section{Mathematical preliminary}

It is well-known how to perform the first two operations of the W sequence 
described above
\cite{MARTINISFock2008}.
We can therefore assume that the circuit was initially prepared in the state 
$|00\dots 001_{r}\ket$,
with only the bus excited. By simultaneously turning on the $N$ couplings,
the W state of the $N$-qubit network can then be generated using a single 
entangling operation (cf.\ \cite{MingYang2004}).

In order to see how this works, we consider 
a formal problem of an ``effective'' Hamiltonian,
\BEqA
  H^{(N+1)} &=& 
g \begin{pmatrix}
		0 &1& 1&1&1&\dots &1\cr
           1& 0 &0 &0&0&\dots &0\cr
           1 &0& 0 &0&0&\dots &0\cr
            1&0&0&0&0&\dots &0 \cr
            1&0&0&0&0&\dots &0\cr
            \dots&\dots &\dots &\dots &\dots &\dots &\dots \cr
            1&0&0&0&0&\dots&0
            \end{pmatrix},
 \EEqA
which operates within a certain $(N+1)$-dimensional Hilbert space
${\cal H}^{(N+1)}$,
 whose physical significance will be clarified below.
 The spectrum of $H^{(N+1)}$ is found to be
 \BEq
 E^{(N+1)} = \mp\sqrt{N}, 0, \dots, 0.
 \EEq
The corresponding eigenvector matrix $S^{(N+1)}$, which
diagonalizes 
$H^{(N+1)}$ via $H^{(N+1)}_{\rm diag}=S^{(N+1)\dagger} H^{(N+1)}S^{(N+1)}$, 
is
given by
\BEq
S^{(N+1)} \sim
\begin{pmatrix}
		-\sqrt{N} &\sqrt{N}& 0&0&0&\dots &0\cr
           1& 1 &-1 &-1&-1&\dots &-1\cr
           1 &1& 1 &0&0&\dots &0\cr
            1&1&0&1&0&\dots &0 \cr
            1&1&0&0&1&\dots &0\cr
            \dots&\dots &\dots &\dots &\dots &\dots &\dots \cr
            1&1&0&0&0&\dots&1
            \end{pmatrix},
 \EEq
where we left the columns of $S^{(N+1)}$ unnormalized for notational simplicity.
 Direct exponentiation then shows that the $N$-dimensional uniform superposition 
 state in this ``effective''
$(N+1)$-dimensional system can be generated via
 \BEq
 \label{eq:uniformstate}
\frac{1}{\sqrt{N}}
\begin{pmatrix}
0\cr
1\cr
1\cr
1\cr
1\cr
\dots \cr
1\cr
\end{pmatrix}
=
i
e^{-iH^{(N+1)}t^{(N)}}
\begin{pmatrix}
1\cr
0\cr
0\cr
0\cr
0\cr
\dots \cr
0\cr
\end{pmatrix},
 \EEq
where 
\BEq
t^{(N)} \equiv \frac{\pi}{2g\sqrt{N}}.
 \EEq

\section{${\rm W}_{N}$ state generation}

Our W state generation scheme is based on the idea that the effective Hamiltonian
$H^{(N+1)}$ considered above should be viewed as operating within the
single-excitation subspace of a network consisting of $N$ qubits coupled to a 
common resonator bus. The corresponding mapping
(extended by linearity)
between the ``effective'' Hilbert space ${\cal H}^{(N+1)}$ and the $(N+1)$-dimensional 
single-excitation subspace of the system
may be chosen to be
\BEq
\begin{pmatrix}
1\cr
0\cr
0\cr
0\cr
0\cr
\dots \cr
0\cr
\end{pmatrix} \rightarrow |00\dots 001_{r}\ket,
\begin{pmatrix}
0\cr
1\cr
0\cr
0\cr
0\cr
\dots \cr
0\cr
\end{pmatrix} \rightarrow |00\dots 010_{r}\ket, \dots .
\EEq
The uniform $N$-dimensional superposition state generated in accordance with
Eq.\ (\ref{eq:uniformstate}) then corresponds to the ${\rm W}_{N}$ state of the 
$N$ qubits attached to the common bus.

In order for this approach to work, the single-excitation subspace has to be well 
isolated from
the rest of the system's Hilbert space. This near-degeneracy condition is typically 
well satisfied in various superconducting qubit architectures whose couplings, 
$g \simeq 100$ MHz, are much smaller than the qubit and resonator level splittings 
of $E_{10}\simeq 10$ GHz.

Let us check that the Hamiltonian $H^{(N+1)}$ arises naturally within the 
single-excitation 
subspace of a capacitively coupled network consisting of superconducting phase qubits 
and a resonator bus.
When projected into the computational subspace spanned by the 
eigenfunctions $|0\ket, |1\ket, |2\ket$ of the individual Josephson phase qubits 
(as well as the resonator), the Hamiltonian of such a network is given by
\BEq
\label{eq:HphybitNetwork}
H
 = \sum_{i=1}^{N}H_{i} + H_{r}
 +\sum_{i=1}^{N}g_{ir}p_{i}p_{r},
\EEq
where the index $i$ numbers the qubits and $r$ labels the bus, with
\BEqA
H_{1} &=&
\begin{pmatrix}
			-E_{10} &0& 0\cr
                0 & 0 &0 \cr
                0 &0& E_{10}-\Delta_{1}
                \end{pmatrix},
\nonumber \\
H_{j} &=&
\begin{pmatrix}
			-E_{10} &0& 0\cr
                0 & \epsilon_{j} &0 \cr
                0 &0& E_{10}+2\epsilon_{j}-\Delta_{j}
                \end{pmatrix}, 
                \quad
                j=2,\dots,N,
\nonumber \\
H_{r} &=&
\begin{pmatrix}
			-E_{r} &0& 0\cr
                0 & \epsilon_{r} &0 \cr
                0 &0& E_{r}+2\epsilon_{r}
                \end{pmatrix}.
\EEqA
The generalized momenta $p_{i}$ and
$p_{r}$ are given by
\BEqA
p_{i} &=&
\lambda_{2} + b\lambda_{5}
 +c\lambda_{7}
=
i\begin{pmatrix}
0 &-1 &        -b_{i}\cr
1& 0 &        -c_{i}\cr
b_{i}& c_{i}& 0
\end{pmatrix},
\nonumber \\
p_{r} &=&
\lambda_{2} +\sqrt{2}\lambda_{7}
=
i\begin{pmatrix}
0 &-1 &     0\cr
1& 0 &     -\sqrt{2}\cr
0& \sqrt{2}& 0
\end{pmatrix},
\EEqA
where $\lambda_{k}$, $k=1,2, \dots, 8$, are the standard 
Gell-Mann generators of the Lie algebra su(3).
In the above, $g_{ir}$ are the qubit-bus coupling constants,
$E_{10}$ is the energy splitting of the first (reference) 
qubit, 
$\epsilon_{j}, \epsilon_{r}$, $j=2,3,\dots,N$, are the energy shifts relative to the 
single-excitation energy of the first qubit, 
$\Delta_{i}$, $i=1,2,\dots,N$, are the qubit anharmonicities, $b_{i}$ and $c_{i}$ 
are the off-diagonal
matrix elements of the $i$th qubit momentum, 
and $E_{r}$ is the resonator energy splitting.
The $(N+1)\times (N+1)$ block
of the Hamiltonian $H$ 
acting within the $(N+1)$-dimensional single-excitation subspace
spanned by $|00\dots 001_{r}\ket, |00\dots 010_{r}\ket, \dots, |10\dots 000_{r}\ket$, 
is then given by the real symmetric matrix,
\BEq
H^{(N+1)} =
\begin{pmatrix}
\epsilon_{r} &g_{Nr}&g_{N-1r}&g_{N-2r}&\dots & g_{3r}&  g_{2r}& g_{1r} \cr
g_{Nr}& \epsilon_{N}& 0& 0& \dots& 0& 0&0\cr
g_{N-1r}& 0& \epsilon_{N-1}& 0&\dots&0&0&0\cr
g_{N-2r}& 0&  0&  \epsilon_{N-2}&\dots&0&0&0\cr
\vdots& \vdots& \vdots & \vdots&\dots&\vdots &\vdots &\vdots\cr
g_{2r}& 0&  0&  0&\dots&0&\epsilon_{2}&0\cr
g_{1r}& 0&  0&  0&\dots&0&0&0
\end{pmatrix}.
\EEq
This immediately shows that the ${\rm W}_{N}$ state can be generated if we 
place all system elements on resonance with each other by choosing
\BEqA
&& \epsilon_{2} = \epsilon_{3} = \epsilon_{4} = \dots = \epsilon_{N} 
= \epsilon_{r} = 0,
\nonumber \\
&& g_{1r} = g_{2r} = g_{3r} = \dots = g_{Nr} = g.
\EEqA

\section{Numerical simulation results}

We tested this scheme on an $N=4$ qubit network
with $g=100$ MHz,
$E_{10}=E_{r}=10$ GHz, $\Delta_{j}=250$ MHz, and
\BEq
p_{i} 
=
i\begin{pmatrix}
0 &-1 &        -0.08\cr
1& 0 &        -1.43\cr
0.08& 1.43& 0
\end{pmatrix}.
\EEq
Assuming the system starts in the excited state $|00\dots 001_{r}\ket$, 
the simulated final state of the system is found to be
\BEq
|{\rm W}_{N}\ket_{\rm sim} =
\begin{pmatrix}
  - 0.0003i\cr
   0.4999  \cr        
   0.4999 \cr
   0.4999 \cr
   0.4999 
   \end{pmatrix},
\EEq
with the corresponding entangling time being $t = 1.2500$ ns.
Ignoring the decoherence effects, the intrinsic fidelity \cite{BENENTIvol2} 
of the found state $|W_{N}\ket_{\rm sim}$ relative to the ideal ${\rm W}_{N}$ state,
is
\BEq
{\cal F}_{N} \equiv |\bra {\rm W}_{N}| {\rm W}_{N}\ket_{\rm sim}|^{2} =   0.9994.
\EEq

\section{${\rm W}_{N+1}$ state generation}

In a similar manner, the ${\rm W}_{N+1}$ state can also be generated, in which the 
resonator is maximally entangled with the qubits. 
This corresponds to the sequence of operations
\BEqA
&&|00\dots 000_{r}\ket \rightarrow |00\dots 010_{r}\ket
\rightarrow
\nonumber \\
&&\frac{|00\dots 001_{r}\ket + |00\dots 010_{r}\ket
+\dots + |10\dots 000_{r}\ket}{\sqrt{N+1}}.
\nonumber \\
\EEqA
In this scenario, we take full advantage of 
qubit tunability to construct the ``effective'' single-excitation Hamiltonian of the form
\BEqA
  H^{(N+1)} &=& 
g \begin{pmatrix}
		2 &1& 1&1&1&\dots &1\cr
           1& 0 &0 &0&0&\dots &0\cr
           1 &0& 0 &0&0&\dots &0\cr
            1&0&0&0&0&\dots &0 \cr
            1&0&0&0&0&\dots &0\cr
            \dots&\dots &\dots &\dots &\dots &\dots &\dots \cr
            1&0&0&0&0&\dots&0
            \end{pmatrix}.
 \EEqA
Its spectrum and the diagonalizing transformation matrix $S^{(N+1)}$ (here shown 
unnormalized)
are
 \BEq
 E^{(N+1)} = 1\mp\sqrt{N+1}, 0, \dots, 0,
 \EEq
and
\BEq
\begin{pmatrix}
		1-\sqrt{N+1} &1+\sqrt{N+1}& 0&0&0&\dots &0\cr
           1& 1 &-1 &-1&-1&\dots &-1\cr
           1 &1& 1 &0&0&\dots &0\cr
            1&1&0&1&0&\dots &0 \cr
            1&1&0&0&1&\dots &0\cr
            \dots&\dots &\dots &\dots &\dots &\dots &\dots \cr
            1&1&0&0&0&\dots&1
            \end{pmatrix},
 \EEq
 respectively. The corresponding W state is generated via
 \BEq
 \label{eq:uniformstate}
\frac{1}{\sqrt{N+1}}
\begin{pmatrix}
1\cr
1\cr
1\cr
1\cr
1\cr
\dots \cr
1\cr
\end{pmatrix}
=
ie^{i\alpha^{(N+1)}}
e^{-iH^{(N+1)}t^{(N+1)}}
\begin{pmatrix}
1\cr
0\cr
0\cr
0\cr
0\cr
\dots \cr
0\cr
\end{pmatrix},
 \EEq
where 
 \BEq
\alpha^{(N+1)} = \frac{\pi}{2\sqrt{N+1}}, 
\quad
t^{(N+1)} \equiv \frac{\pi}{2g\sqrt{N+1}},
 \EEq
 provided we detune the qubits from the resonator by $2g$,
\BEqA
&& \epsilon_{2} = \epsilon_{3} =  \epsilon_{4} = \dots = \epsilon_{N} = 0,
\quad
\epsilon_{r} = 2g, \nonumber \\
&& g_{1r} = g_{2r} = g_{3r} = \dots = g_{Nr} = g.
\EEqA
The duration of the entangling pulse is now $t = 1.1180$ ns, with the
simulated single-excitation final state of the network being
\BEq
|{\rm W}_{N+1}\ket_{\rm sim} =
\begin{pmatrix}
   0.4471 - 0.0003i\cr
   0.4472\cr
   0.4472 \cr
   0.4472\cr
   0.4472 
   \end{pmatrix},
\EEq
with fidelity 
\BEq
{\cal F}_{N+1} =   0.9997.
\EEq







\begin{thebibliography}{}


\bibitem{MingYang2004}
Ming Yang, You-Ming Yi, Zhuo-Liang Cao,
Int. J. Quantum Inf. {\bf 2}, 231 (2004).

\bibitem{MIGLIORE2006}
R. Migliore, K. Yuasa, H. Nakazato, and A. Messina, 
Phys. Rev. B {\bf 74}, 104503 (2006).

\bibitem{MARTINISFock2008}
M. Hofheinz, E. M. Weig, M. Ansmann, R. C. Bialczak, E. Lucero, M. Neeley, 
A. D. O'Connell, H. Wang, J. M. Martinis, and A. N. Cleland, Nature {\bf 454}, 
310 (2008).

\bibitem{BENENTIvol2}
G. Benenti, G. Casti, G. Strini, {\it Principles of Quantum Computation and Information},
Vol. II, World Scientific (2007).
































































\end{thebibliography}
\end{document}